\documentclass[preprint2]{aastex}

\shorttitle{The environments of SNe Ia} \shortauthors{Meng et al.}

\begin{document}


\title{The environments of Type Ia supernovae with different relative equivalent width of Si II feature in their spectra}


\author{Xiang-Cun Meng$^{\rm 1,2}$, Ju-Jia Zhang$^{\rm 1,2}$, Xu-Lin Zhao$^{\rm 3}$, Li-Ping Li$^{\rm 1,4}$, Xiao-Feng Wang$^{\rm 5}$ }
\affil{$^{\rm 1}$Yunnan Observatories, Chinese Academy of Sciences, 650216 Kunming, PR China\\
$^{\rm 2}$ Key Laboratory for the Structure and Evolution of
Celestial Objects, Chinese Academy of Sciences, 650216 Kunming, PR
China\\
$^{3}$School of Science, Tianjin University of Technology, Tianjin 300384, China\\
$^{4}$University of Chinese Academy of Sciences, Beijing 100049, China\\
$^{5}$Physics Department and Tsinghua Center for Astrophysics
(THCA), Tsinghua University, Beijing 100084, China}
\email{xiangcunmeng@ynao.ac.cn}





\begin{abstract}
Although type Ia supernovae are so important in many astrophysical field, e.g. in cosmology, their explosion mechanism and progenitor system are still unclear. In physics, the relative equivalent width (REW) of the
Si II 635.5 nm absorption feature reflects the velocity interval of silicon in the supernova ejecta and then may provide constraints on the explosion mechanism of SNe Ia.  In this paper, we divide the SNe Ia into broad line (BL) and normal line (NL) subsamples based on their REW of  Si II 635.5 nm absorption lines around maximum light, and find that the BL SNe Ia have a dimmer mean brightness than NL ones, which possibly results from their different metallicities. However, based on the pixel statistics study on the environments of two subsamples, we do not find any significant potential difference on the environments between BL and NL SNe Ia, which implies that the explosion mechanism of SNe Ia could be independent of their progenitor populations.
\end{abstract}





\section{INTRODUCTION}\label{sect:1}
Taking type Ia supernovae (SNe Ia) as standard candles to measure
cosmological parameters, it is found that the expansion of the
Universe is accelerating, which means that the Universe is
dominated by a mysterious dark energy (\citealt{RIE98};
\citealt{PER99}). Now, SNe Ia are used to measure the equation of
dark energy (see \citealt{MENGXC15} and references in it). On SNe
Ia, a consensus has bee achieved that a SN Ia is derived from the
thermonuclear explosion of a carbon-oxygen white dwarf (CO WD) in
a binary system (\citealt{HF60}), but as far as the explosion
physics and the evolution of the WD towards explosion are
concerned, the arguments are endless, i.e. various explosion
models and progenitor models are proposed (\citealt{HN00};
\citealt{GOOBAR11}; \citealt{WANGB12}; \citealt{HILLEBRANDT13};
\citealt{MAOZ14}; \citealt{JHA19}).  According to the nature of the companion star of the WD, two competing progenitor scenarios have been
proposed: the single degenerate (SD) model, where the
companion is a nondegenerate star, e.g. a main-sequence (MS), a red-giant (RG) or a helium star (\citealt*{WI73}; \citealt{NTY84}), and the double-degenerate (DD) model, which involves the merger of two CO WDs with a total mass of larger than the Chandrasekhar mass limit (M$_{\rm Ch}$) (\citealt{IT84}; \citealt{WEB84}). In addition to the above models, the double-detonation model is also frequently discussed, where the CO WD companion is a helium WD or a helium star (\citealt{LIVNE90}; \citealt*{WOOSLEY94}; \citealt{HOEFLICH96}; \citealt*{SHENK14}). At present, which model dominates the production of SNe Ia is still a key subject of debate (\citealt{JHA19}; \citealt{RUIZLAPUENTE19}).  Based on the mass of the CO WD at the moment of supernova explosion and the propagating flame style,  sub-M$_{\rm Ch}$,  M$_{\rm Ch}$ and super-M$_{\rm Ch}$ explosion models are proposed (\citealt{HN00};  \citealt{HILLEBRANDT13}).

Strictly speaking, the research on the explosion model and progenitor
model should not be independent, e.g. the initial conditions of the
explosion of a WD should be determined by the evolution of its
progenitor, but they are generally treated as free parameters. Then,
how to connect the explosion model and the progenitor model
together is becoming an interesting subjects. On one side, the
spectra of SNe Ia may reflect the explosion mechanism to some
extents although different explosion models could predict quiet
similar spectra (\citealt{WANGLF08}; \citealt{HILLEBRANDT13}). On
the other hand, the population of SNe Ia may partly reflect the
nature of the progenitor system of SNe Ia (\citealt{WANGXF13}).
Therefore, finding a connection between spectral
feature and the population of SNe Ia has a potential to link the
explosion model and progenitor model of SNe Ia.

It is widely known that SNe Ia are not a homogeneous population for their photometric and spectroscopic diversities (\citealt{TAUBENBERGER17}). In 1991,  two peculiar SNe Ia were
discovered, i.e.  SN 1991bg and 1991T. 1991bg is fainter than normal SNe Ia by about 2 magnitudes (\citealt{FIL92a}; \citealt{LEIBUNDGUT93}), while SN 1991T is brighter than normal ones by 0.4 magnitudes in the V-band (\citealt{FIL92b}; \citealt{PHILLIPS92}). Since then, SNe Ia was classified into different subclasses based on different measurable properties of SNe Ia (\citealt{BENETTI05};
\citealt{BRANCH09}; \citealt{WANGXF09}). Such classifications may
be helpful to diagnose different progenitor populations and/or different explosion mechanisms that different subclasses are derived from. For example, based on the different ejecta
velocities of SNe Ia, \citet{WANGXF13} found that normal SNe Ia are
derived from two different populations, i.e compared with those
with normal-velocity (NV; $v_{\rm Si, max}\le12000\,{\rm km\,s^{\rm -1}}$) ejecta, SNe Ia with high-velocity (HV; $v_{\rm Si, max}\ge12000\,{\rm km\,s^{\rm -1}}$) ejecta tend to occur in the inner and brighter regions of their host galaxies, i.e. HV SNe Ia favor metal-rich progenitor systems.  One of the reasons that supports the existence of the different subclasses is that the distribution of the parameters that describe the
properties of SNe Ia may not be simply described by one Gaussian
distribution, even though only the so-called normal SNe Ia are
considered (\citealt{WANGXF13}). For an example, the distributions
of the parameters describing the maximum absolute brightnessof SNe
Ia has a peak with a long faint tail and a sharp cut at luminous
end (\citealt{HICKEN09}; \citealt{LIWD11b}; \citealt{BLONDIN12};
\citealt{ASHALL16}). However, do SNe Ia really need so many
subclasses since more and more transitional events are discovered, suggesting a continuous distribution from super-luminous to normal, and to sub-luminous SNe Ia (\citealt{PARRENT14}; \citealt{ZHANGJJ17}; \citealt{LIZT22}). In particular, many works argued that 1991bg-like SNe may not be a real physically
distinct subgroup (\citealt{BRANCH09}; \citealt{DOULL11};
\citealt{HERINGER17}), and there even exist some SNe Ia, which
should belong to normal SNe Ia although they have a very large
$\Delta m_{\rm 15}(B)$ (\citealt{GALL17}). Is there such a parameter which may
describe all SNe Ia and whose distribution may be
fitted by a simple Gaussian? 

Firstly, we tried to find a parameter based on the following two
criteria: 1) the parameter may reflect the characteristics of the
explosion of SNe Ia, and 2) its distribution is a simple Gaussian.
\citet{MENGXC17} (Paper I) found a correlation between the
polarization and relative equivalent width (REW) of the Si II
635.5 nm absorption feature at -5 days after the maximum light,
where the polarization may provide the information of the chemical
structure of the supernova ejecta and then constrain the explosion
model, and the REW reflect the distribution of silicon in
supernova ejecta. Actually, the REW, which is defined as the ratio
of the pseudo equivalent width (pEW) to the relative depth ($a$)
of the Si II 635.5 nm absorption feature (\citealt{SILVERMAN12}),
reflects the velocity difference between inner and outer boundary
of silicon in supernova ejecta, and then the explosion characteristics to some extent (\citealt{MENGXC17}). The parameter, REW, fulfills our two criteria. In this paper, we want to check whether or not SNe Ia with different REWs
originates from different populations.

In section \ref{sect:2}, we describe our method, and present the
calculation results in section \ref{sect:3}. We show discussions
and our main conclusions in section \ref{sect:4}.

\begin{figure}
	\centerline{\includegraphics[angle=0,scale=.55]{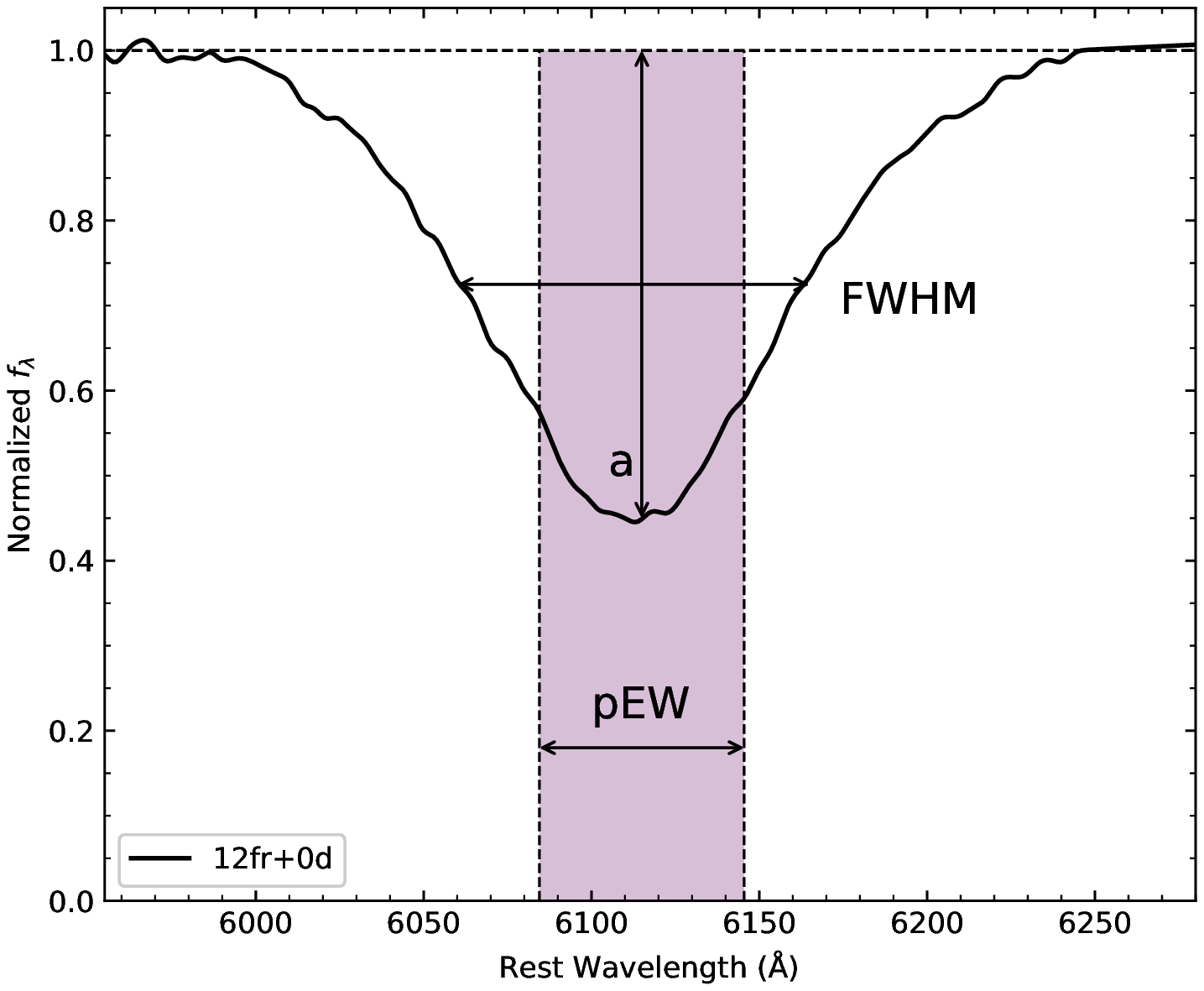}}
	\caption{Schematic diagram illustrating the measurement of pEW and $a$ of the Si II 635.5 nm absorption feature of SNe Ia from a spectrum of SN 2012fr at maximum	brightness (\citealt{ZHANGJJ14}), in addition to its full width at half maximum.}\label{pewa}
\end{figure}

\section{DATA}
\label{sect:2}

As described in \citet {SILVERMAN12}, the measurement of pEW and $a$ is illustrated in Fig.~\ref{pewa}. For a given absorption line in the spectra of a SN Ia, different line profiles may lead to the same  pEW. To eliminate the effect of the different line profiles, \citet{MENGXC17}  introduce the REW. The definition and the  measurement of the REW value here are the same to that in Paper I, i.e.
\begin{equation}
	{\rm REW}=\frac{\rm pEW}{a}.\label{eq:1}
\end{equation}
Following Paper I, we measured the REW value of the Si II 635.5 nm
absorption feature of SNe Ia around maximum light, which is the
most significant feature in the optical spectra around the maximum
light and almost not blend with other features in the optical
spectra.  All the data about SNe Ia used in this paper have been published. The SN Ia sample is mainly from \citet {SILVERMAN12} and \citet{WANGXF13}, and there was
no bias in the collection of the sample (see also Paper I). Some parameters of
SNe Ia, e.g. $\Delta m_{\rm 15}{\rm (B)}$ and intrinsic B - V color,  are from \citet{BLONDIN12}. In Paper I, we found that the
intrinsic distribution of the REW around maximum light may be
fitted by a Gaussian with an average value of 159.7 ${\rm \AA}$
and $\sigma=45.56~{\rm \AA}$. Here, we do not classify SNe Ia as people usually did (\citealt{BRANCH93}).
We divide SNe Ia into two subclasses based on the REW value, i,e.
broad line (BL) one with ${\rm REW}>159.7~{\rm \AA}$ and normal
line (NL) one with ${\rm REW}<159.7~{\rm \AA}$. If BL and NL SNe
Ia share the same origin, they would occur in similar stellar
environments, and vice versa. Then, we try to investigate whether or not there are any
signatures that the different subclasses originate from different
populations following the method in \citet{WANGXF13} and \citet{ANDERSON15,ANDERSON15b}. The stellar
environments at the location of supernova explosion may reflect the
information of the populations of SNe Ia, and can be tested by
checking SN positions in their host galaxies and the properties of
the host galaxies. The location information of SNe Ia in their host galaxies is mainly from  \citet*{WANGXF13}, \citet{ANDERSON15b} and \citet{KARAPETYAN22}. Some of the parameters of the SN host galaxies are from \citet*{WANGXF13} and the others are
obtained from two online astronomical databases: \citet{NED}\footnote{http://ned.ipac.caltech.edu/} and
HyperLeda\footnote{http://leda.univ-lyon1.fr/}(\citealt{MAKAROV14}).




\begin{figure}
\centerline{\includegraphics[angle=270,scale=.35]{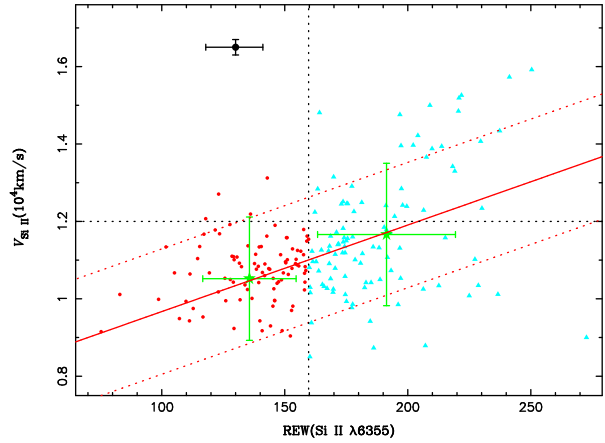}}
\caption{Relations between the photospheric velocity and REW of Si II 635.5 nm
absorption feature around maximum light. Blue and red points represent BL and NL SNe Ia, respectively. Two green crosses show the average values with $1\sigma$ statistical error  for BL and NL SNe Ia, respectively. The red solid line is the linear fit, and the two red
dotted lines show the $1\sigma$ region of the fit. The two black
dotted lines show the boundaries between NV and HV, and BL and NL
SNe Ia, respectively. The black cross at the left-upper region
shows the typical error of the data points. The data are from \citet {SILVERMAN12} and \citet{WANGXF13}.}\label{rrva}
\end{figure}

\section{RESULT}\label{sect:3}
\subsection{Relations between the photospheric velocity and REW}\label{sect:3.1}
In \citet{WANGXF13}, SNe Ia are divided into NV
and HV groups based the photospheric expansion velocity of supernova ejecta, where 12000 km/s is the velocity boundary. In Fig.~\ref{rrva}, we show the relation between the
velocity and REW of Si II 635.5 nm absorption feature around
maximum light. As found in \citet{WANGXF13}, most of SNe Ia belong
to NV SNe Ia, and most of SNe Ia have a photospheric velocity
between 10000 and 12000 km/s. Generally, there is trend that the
higher the photospheric velocity, the higher the REW value, as shown by
the linear fit in Fig.~\ref{rrva}, although the scatter is still large. However, the subclasses based on different photospheric velocities and on the REW do not correspond
one to one, e.g. only a small part of BL SNe Ia belong to HV SNe Ia,
and most of them belong to NV group, which results in a small
difference of the average photospheric
velocity between BL and NL SNe Ia. Nevertheless, almost all HV SNe Ia tend to belong to BL SNe Ia and at the same time, almost all NL SNe Ia belong to NV ones.

\begin{figure*}
\centerline{\includegraphics[angle=270,scale=.45]{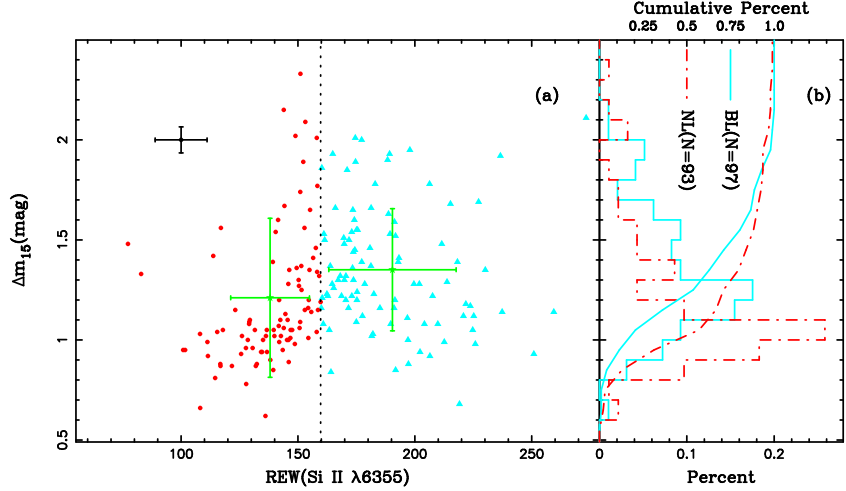}}
\caption{a) The distribution of SNe Ia in the plane of $\Delta m_{\rm 15}{\rm (B)}$ and
REW. The two green crosses show the average value with $1\sigma$ statistical error for NL and BL SNe Ia, respectively, and the black cross presents the typical
observational error of the points. The vertical dotted line shows
the boundary of NL (red) and BL (blue) SNe Ia. b) The histogram
and cumulative percent of $\Delta m_{\rm 15}{\rm (B)}$ for NL and
BL SNe Ia, respectively. The data are from \citet{BLONDIN12} and \citet{SILVERMAN12}.}\label{rrdeta}
\end{figure*}

\begin{figure*}
\centerline{\includegraphics[angle=270,scale=.45]{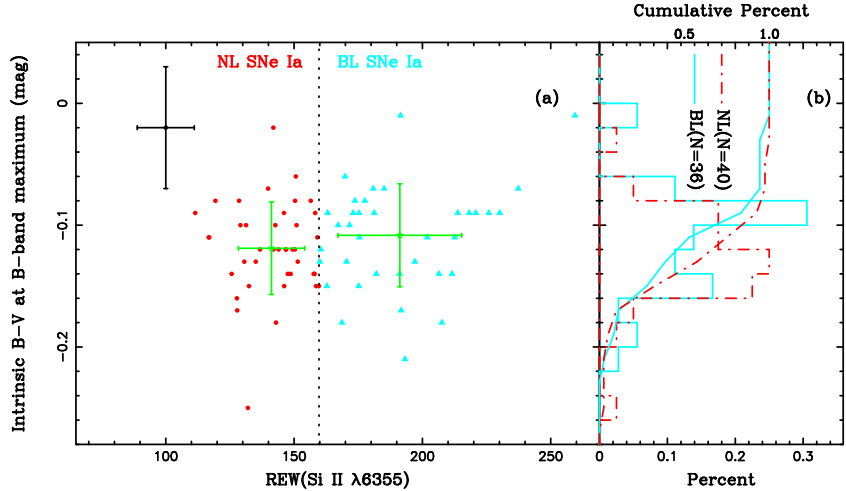}}
\caption{a) The distribution of SNe Ia in the plane of intrinsic (B-V) color and REW at B-band maximum brightness. The
two green crosses show the average value with $1\sigma$ statistical error for NL and BL SNe Ia, respectively, and the black cross presents the typical
observational error of the points. The vertical dotted line shows
the boundary of NL (red) and BL (blue) SNe Ia. b) The histogram
and cumulative percent of intrinsic (B-V) color at B-band maximum
for NL and BL SNe Ia, respectively. The data are from \citet{BLONDIN12} and \citet{SILVERMAN12}.}\label{rrcolor}
\end{figure*}

\subsection{$\Delta m_{\rm 15}{\rm (B)}$ and REW}\label{sect:3.2}
$\Delta m_{\rm 15}{\rm (B)}$, the B band magnitude drop during the
first 15 days following the maximum brightness, is usually taken as
the measurement of the maximum absolute brightness of SNe Ia, i.e.
the larger the value of $\Delta m_{\rm 15}{\rm (B)}$, the dimmer
the maximum absolute brightness (\citealt{PHILLIPS93}). However, the
distribution of $\Delta m_{\rm 15}{\rm (B)}$ can not be fitted by a Gaussian as REW (\citealt{HICKEN09}; \citealt{BLONDIN12}). In panel (a) of
Fig.~\ref{rrdeta}, we show the distribution of SNe Ia in the plane of $\Delta m_{\rm 15}{\rm (B)}$ and REW. On the point of average, the BL SNe Ia have
a larger average value of $\Delta m_{\rm 15}{\rm (B)}$ than the NL
SNe Ia, which means that the BL SNe Ia are dimmer than the NL ones
on average although the dimmest SNe Ia belong to NL SNe Ia. The
mean $\Delta m_{\rm 15}{\rm (B)}$ for BL and NL SNe Ia are
$1.35\pm0.31$ mag and $1.21\pm0.40$ mag, respectively. The
distribution of $\Delta m_{\rm 15}{\rm (B)}$ for BL and NL SNe Ia
also show such a trend [panel (b) of Fig.~\ref{rrdeta}]. The
distribution of BL SNe Ia peaks at $\Delta m_{\rm 15}{\rm
(B)}=1.25$ mag, a value larger than that of NL SNe Ia by 0.2 mag.
A Kolmogorov-Smirnov (K-S) test only gives a probability of
$4.2\times10^{\rm -6}$ that BL and NL SNe Ia come from the same
mother sample for the $\Delta m_{\rm 15}{\rm (B)}$ distributions. Here, we must emphasize that we do not classify SNe Ia as people usual did. Roughly, 60\%  of 1991bg-like SNe belong to BL and 40\% are to NL one. At the same time, about 20\% of
1991T-like SNe belong to BL SNe Ia and 80\% are NL one. Such a
distribution of 1991bg- and 1991T-like SNe partly contributes to
the difference of the mean $\Delta m_{\rm 15}{\rm (B)}$ between BL
and NL SNe Ia, but may not explain their different peak positions
of the $\Delta m_{\rm 15}{\rm (B)}$ distributions since the peak
positions are for normal SNe Ia.  We discuss the possible origin of the different average brightness between BL and NL SNe Ia in Sec.~\ref{sect:4}.

Similarly, the intrinsic (B-V) color at the B-band maximum brightness for BL SNe Ia is on average redder than that for NL ones by $\sim0.01$ mag (see Fig.~\ref{rrcolor}),
which is consistent with previous discovery that intrinsically dim
SN Ia are redder (\citealt{RIESS96}; \citealt{BRANCH98}). However,
the observational error of the intrinsic (B-V) color is much
larger than the difference of the average color between BL and NL
SNe, and then the distributions of the intrinsic (B-V) color are
not significantly different between BL and NL SNe, i.e. a K-S test
shows a probability of 27\% that the distributions for BL and NL
SNe are from the same mother sample.

Generally, the $^{\rm 56}$Ni production during the explosion
dominates the maximum luminosity of SNe Ia (\citealt{ARNETT82}),
but no consensus has been achieved for the origin of the variation
in the $^{\rm 56}$Ni production of different SNe Ia
(\citealt{PODSIADLOWSKI06}). In observations, the age, metallicity
of the population as well as the mass of their host galaxy may
affect the maximum brightness of SNe Ia (see the review of
\citealt{MENGXC15}). However, whether or not the difference of the
brightness between NL and BL SNe Ia originates from different
populations is still needed to be carefully investigated. To this end,
we will show whether or not there is a difference between BL and SNe Ia  for the characteristics representing the population of SNe Ia in the following sections.

\begin{figure*}
\centerline{\includegraphics[angle=270,scale=.45]{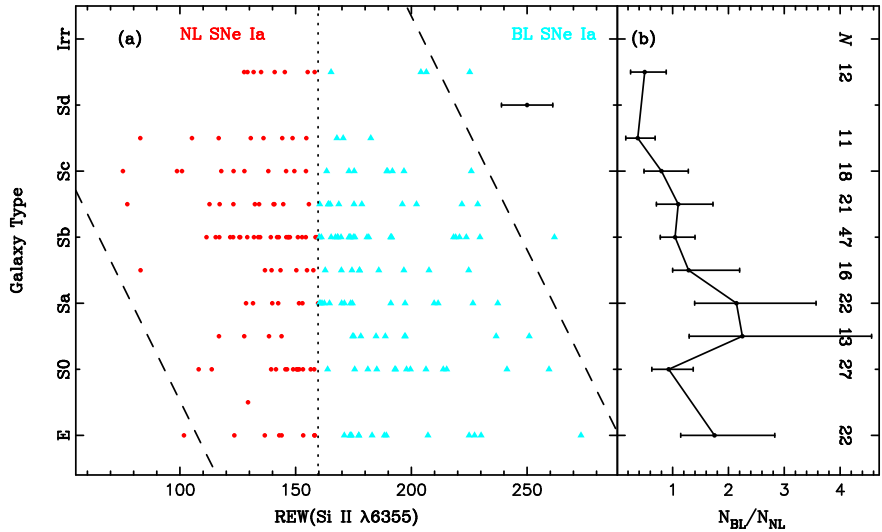}}
\caption{a) The Hubble type of host galaxy vs. the REW of SNe Ia around the maximum brightness. The two dashed lines show the main region for SNe Ia in the plane and the
vertical dotted line shows the boundary of NL (red) and BL (blue)
SNe Ia. b) The number ratio of BL to NL SNe Ia, and the error bars are calculated  by
assuming a binomial distribution (\citealt{CAMERON11}). The numbers
show the number of SNe Ia in every galaxy type. The data are from \citet{SILVERMAN12}.}\label{rrhost}
\end{figure*}

\subsection{The host galaxy}\label{sect:3.3}
Firstly, the type of the host galaxy of SNe Ia may reflect the population of SNe Ia to
some extent. For example, it is well known that the most luminous
SNe Ia are always discovered in spiral galaxies, while both spiral and elliptical galaxies may host dimmer SNe Ia, which results in a dimmer mean peak SN brightness in elliptical than in spiral ones (\citealt{HAMUY96}; \citealt{BRANDT10}). For galaxies, there are  some systematic trends for
the stellar population along the Hubble sequence, i.e. the early
the galaxy type, the older the stellar population and the higher
the metallicity (\citealt{ROBERTS94}; \citealt{KENNICUTT98}). In panel (a) of
Fig.~\ref{rrhost}, we show the Hubble type of
host galaxy vs. the REW of SNe Ia around the maximum brightness and in panel (b) of Fig.~\ref{rrhost}, we also show the  number ratio of BL to NL SNe Ia in different type galaxies. It is clearly shown in
panel (a) that compared with the NL ones, the BL SNe Ia
tend to occur in relative early type Galaxy.  According, we find a trend that, whatever in spiral or early type galaxies, the earlier the galaxy type, the
higher the number ratio of BL to NL SNe Ia. These facts are consistent with the results that on average, the BL SNe Ia are dimmer than NL ones in section \ref{sect:3.2}.

\begin{figure*}
\centerline{\includegraphics[angle=270,scale=.45]{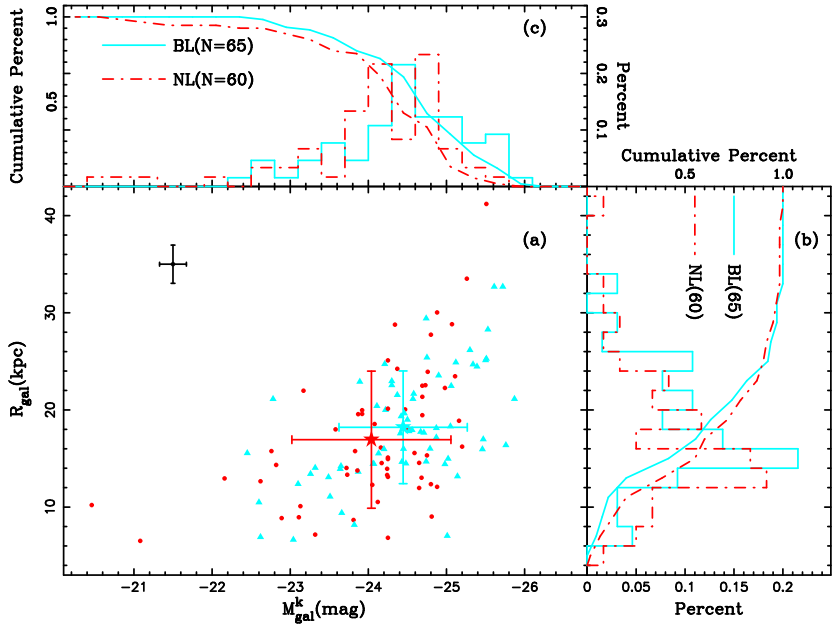}}
\caption{A comparison of the physical sizes and K-band absolute
magnitude distributions of the host galaxies for BL and NL SNe Ia,
respectively, where the data is from \citet{WANGXF13}. The plot is
projected onto the two side panels where a histogram and
cumulative percents are shown for the dimensions and absolute
magnitude of the host galaxies.}\label{rgmk}
\end{figure*}

In Fig.~\ref{rgmk}, we show the comparison of the physical sizes,
$R_{\rm gal}$, and K-band absolute magnitude distributions of the
host galaxies for BL and NL SNe Ia. Here, the physical sizes of SN
host galaxies are represented by the B-band light radius at 25 mag
arcsec$^{\rm -2}$ isophote (see \citealt{WANGXF13}). In the figure, there are small difference for the
sizes and K-band absolute magnitude of the host galaxies between
NL and BL SNe Ia. On average, the BL SNe Ia tend to occur in
slightly larger and brighter host galaxies than the NL SNe Ia. 
The average radius of host galaxies for BL and NL SNe Ia are $18.21\pm5.8$
kpc and $16.93\pm7.1$ kpc, respectively and the mean absolute
magnitude of the galaxies for BL and NL SNe Ia are $-24.45\pm0.82$
mag and $-24.04\pm1.02$ mag, respectively. However, the statistical errors are so large that the distributions of the radius and the absolute magnitude of host
galaxies for BL and NL SNe Ia are not significantly different. A
two-dimensional K-S test shows that there is a probability of
15.9\% that the two subclasses come from the same mother sample, i.e. the distributions are indistinguishable.

\begin{figure*}
centerline{\includegraphics[angle=270,scale=.45]{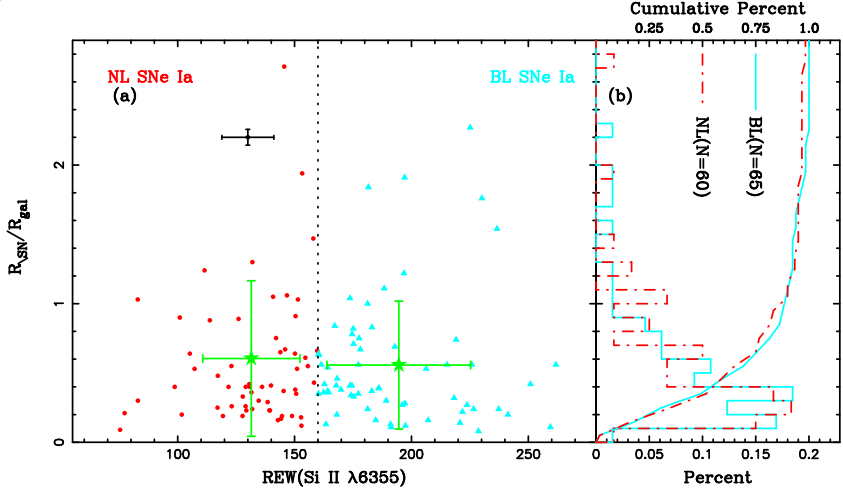}}
\caption{a) The fractional radial distance of SNe Ia in their
the host galaxies vs. the REW value of the SNe Ia around the maximum brightness. The green crosses
show the average values of $R_{\rm SN}/R_{\rm gal}$ and REW, and
the lengthes of the crosses represent the 1$\sigma$ statistic
error for BL and NL SNe Ia, respectively. The black cross shows
the typical observational error for $R_{\rm SN}/R_{\rm gal}$ and
REW. b) The histogram and cumulative percent of $R_{\rm SN}/R_{\rm
gal}$ for NL and BL SNe Ia, respectively.  The data are from \citet{SILVERMAN12} and \citet{WANGXF13}.}\label{rrgal}
\end{figure*}

\subsection{The fractional radial distance in host galaxies}\label{sect:3.4}
Generally, the global parameters of the
host galaxy of a SN Ia are not a good tracer of the stellar population of the SN Ia at the
site of a supernova explosion since the global parameters may  just reflect the average information of the stellar
population in the host galaxy, and could erase the intrinsic
information of the stellar population of the SN Ia. The most direct method constraining the progenitor nature of
a SN Ia is to investigate its pre-explosion image, but such cases are too rare to obtain a statistical meaning (\citealt{LIWD11a}; \citealt{MCCULLY14}). A statistical
analysis of a fractional flux or a fractional radial distance is a viable method to investigate the environments at the position of a SN Ia  in its host galaxy and then constrain its progenitor nature (\citealt{FRUCHTER06}; \citealt*{WANGXF13}; \citealt{ANDERSON15}; \citealt{MENGXC19}).

Following \citet{WANGXF13}, we use the ratio of $R_{\rm SN}$ to $R_{\rm gal}$ to represent the fractional radial distance of a SN Ia in its host galaxy, where
the $R_{\rm SN}$ is the radial distance of the SN Ia from its host
nucleus (see also \citealt{HAKOBYAN16}). Then, we  investigate whether or not there is any bias for the distribution of $R_{\rm SN}/R_{\rm	gal}$ between BL and NL SNe Ia. Fig. \ref{rrgal} shows the fractional radial distance of a SN Ia in its host galaxy vs. the REW of the SN Ia. It is clearly shown in
the figure that the distributions of the $R_{\rm SN}/R_{\rm
gal}$ for BL and NL SNe Ia are quite similar, i.e. a similar peak
position and almost the same cumulative percent of $R_{\rm
SN}/R_{\rm gal}$. In addition, whatever for BL or NL SNe Ia, it is
difficult to find them in the inmost region of their host
galaxies, which is an observational selection bias against SN
discovery (\citealt{CAPPELLARO97,CAPPELLARO99}). A K-S test finds a
probability of 87.1\% that BL and NL SNe Ia have the same radial
distribution in their host galaxies. Such a high probability means
that there is not difference for the distribution of 
$R_{\rm SN}/R_{\rm gal}$ between BL and NL SNe Ia, i.e. BL and NL SNe Ia
originate from the same stellar population.

\begin{figure*}
\centerline{\includegraphics[angle=270,scale=.45]{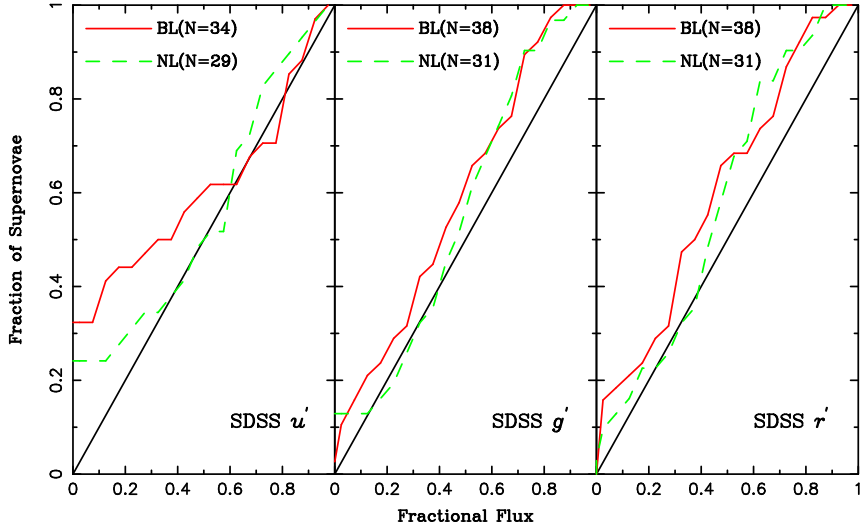}}
\caption{The cumulative distributions of the fractional flux of the
host galaxies at the SN explosion site for BL and NL SNe Ia,
respectively, in $u^{'}$, $g^{'}$ and $r^{'}$ bands. The data are from \citet{WANGXF13}.}\label{fdis}
\end{figure*}

\subsection{The fractional flux}\label{sect:3.5}
The fractional flux of a SN Ia represents the fraction of total host light in pixels fainter than or equal to the light in the pixel of the SN Ia site in its host-galaxy image. A statistical analysis of the fractional flux of the host galaxies
at the SN explosions may reflect the populations of SNe Ia in
their host galaxies, and is an additional constrains on the
progenitors of SNe Ia besides the radial distance
(\citealt{FRUCHTER06}). Generally, the
core-collapse SNe approximately linearly trace the star-formation (light)
region in their host galaxies, while SNe Ia do not
(\citealt{ANDERSON15}). In Fig.~\ref{fdis}, we show the cumulative
distributions of the fractional flux of the host galaxies at the SN
explosion site  in the Sloan Digital Sky Survey (SDSS) $u^{\rm '}$ , $g^{\rm '}$,
and $r^{\rm '}$ bands for BL and NL SNe Ia,
respectively, as in \citet{WANGXF13}.  Generally, a young population traces the diagonal
line, as core-collapse SNe do, while an old population lies far away from the
line in the plot (\citealt{ANDERSON15}). Clearly,  the distributions of the fractional flux for BL and NL SNe Ia are quite similar with each
other, whatever the observational band is. In other words, the
BL and NL SNe Ia track their hosts' light as the same pattern,
since K-S tests give very high probabilities that the fractional
fluxes for BL and NL SNe Ia are from the same mother sample
($P=71.2\%$ in $u^{'}$, 93.8\% in $g^{'}$ and 79.7\% in $r^{'}$
band, respectively). In addition, as found by \citet{ANDERSON15b},
SNe Ia do not trace $u^{'}$ light, compared with the $g^{'}$ and
$r^{'}$ bands (with a probabilities of $P=2\%$ in $u^{'}$ band), which means
that SNe Ia do not trace the star-formation region in their host
galaxies as CC SNe do. However,  SNe Ia well trace $g^{'}$ and $r^{'}$ band lights (with a probabilities of 42\% and 50\% in $g^{'}$ and $r^{'}$ bands, respectively), which indicates that the $g^{'}$ and $r^{'}$ band lights would not be good tracers for the star
formation in a galaxy. On the contrary,  the NUV and H$\alpha$ bands are
good choices for tracing the star formation (\citealp{ANDERSON15,ANDERSON15b}).

\begin{figure*}
	\centerline{\includegraphics[angle=270,scale=.45]{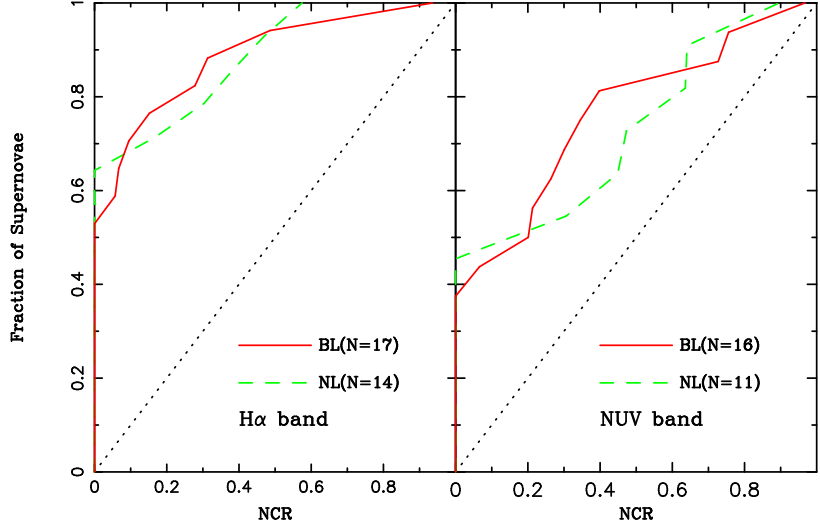}}
	\caption{The cumulative NCR distributions of BL (red solid) and NL (green dashed) SNe Ia in H{$\alpha$} and NUV bands, respectively. The NCR data of the host galaxies at the SN Ia explosion	site are from \citet{ANDERSON15b}.}\label{ncrdis}
\end{figure*}

\subsection{The NCR pixel value function }\label{sect:3.6}
The statistical analysis of  a normalized cumulative rank
(NCR) pixel value function of the host galaxies at the SN
explosion site is another viable method to investigate the population of SNe Ia.  The NCR value of a pixel is equal to the flux-count ratio between this pixel and the pixel with the highest flux count within the image of a host galaxy (\citealt{ANDERSON08}). Generally,  the closer the cumulative line of the NCR line for a SN sample to the diagonal line, the younger the population of  the sample (\citealt{ANDERSON08,ANDERSON15b}). In Fig.~\ref{ncrdis}, we show the cumulative NCR distributions for BL  and NL SNe Ia in H{$\alpha$} and NUV bands, respectively. It is clearly shown in the figure that cumulative NCR distributions  do not trace the diagonal line, which indicates that the progenitors of SNe Ia are not
as young as CC SNe (see also \citealt{ANDERSON15,ANDERSON15b}). Nevertheless, no matter what the band is, the cumulative NCR distributions are quite similar between BL  and NL SNe Ia. K-S tests for the NCR
distributions  between the NL and BL SNe Ia show that the probability that the two
subsamples are from the same mother sample are 99.96\% and 66.9\% for the H$\alpha$ and NUV bands, respectively, i.e. they arise from the same mother sample.

\begin{figure*}
	\centerline{\includegraphics[angle=270,scale=.45]{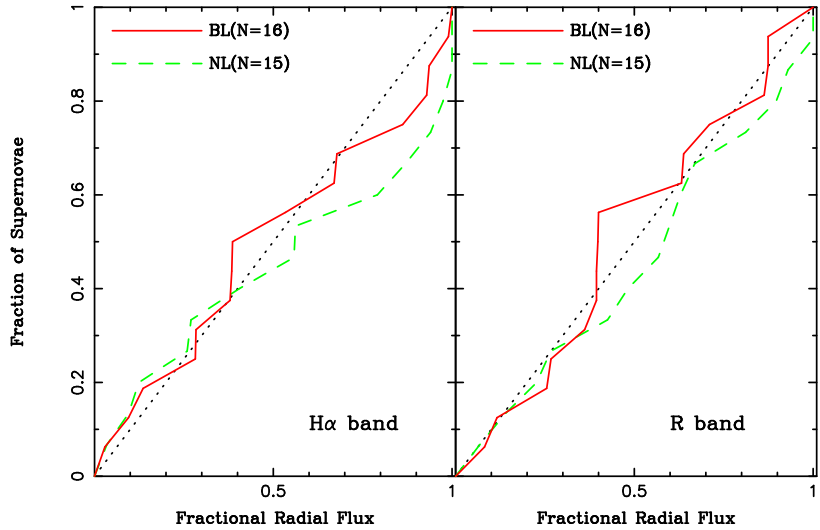}}
	\caption{The cumulative distributions of the Fr value for the BL (red solid) and NL (green dashed) SNe Ia in H{$\alpha$} (left) and R (right) bands, respectively. The Fr data of the host galaxies at the SN Ia explosion site are from \citet{ANDERSON15b}.}\label{frdis}
\end{figure*}

\subsection{Radial analysis}\label{sect:3.7}
It is believed that a SN Ia is much older than a CC SN and then the explosion site of the SN Ia is far away from its birth site. Generally, stellar populations with different ages and/or metallicities locate at different
characteristic galactocentric radial positions  in a galaxy, and then one may obtain the information of the population (or progenitor) of SNe Ia by exploring a `Fr' fractional flux value at the positions of the SNe Ia in their host galaxies. Here, following  \citet{ANDERSON15b}, H{$\alpha$}  and R bands are considered.  Based on the definition of Fr,  Fr = 0 for a SN Ia indicates that the SN Ia locates at the central peak pixel in the H{$\alpha$} or R-band image of its host galaxy, while
Fr = 1 means that the SN Ia explodes in the outer region of its
host galaxy, where the H{$\alpha$} or R-band flux is even equal to the sky
value  (please see \citealt{ANDERSON15b} for the definition of Fr in details). In other words, a low value of Fr means a relatively young population.  In Fig.~\ref{frdis}, we show the cumulative distribution of the Fr value for the BL and NL SNe Ia in H{$\alpha$} and R bands, respectively.  Again, the distributions are similar with each other, no matter what the band is. A two-dimensional K-S test gives a possibility of 61.1\% that  two subsamples are from the same mother sample. In addition,  all the host galaxies in \citet{ANDERSON15b} are star-forming galaxies, which indicates the more SNe Ia in the sample of \citet{ANDERSON15b} belong to young population (see also \citealt{MENGXC19}). However, the number ratio of BL to NL is roughly  $\sim1$ in the sample of \citet{ANDERSON15b}, which may be taken as another piece of evidence  that the classification for BL and NL SNe Ia is independent of their stellar population.  

\begin{figure*}
	\centerline{\includegraphics[angle=270,scale=.45]{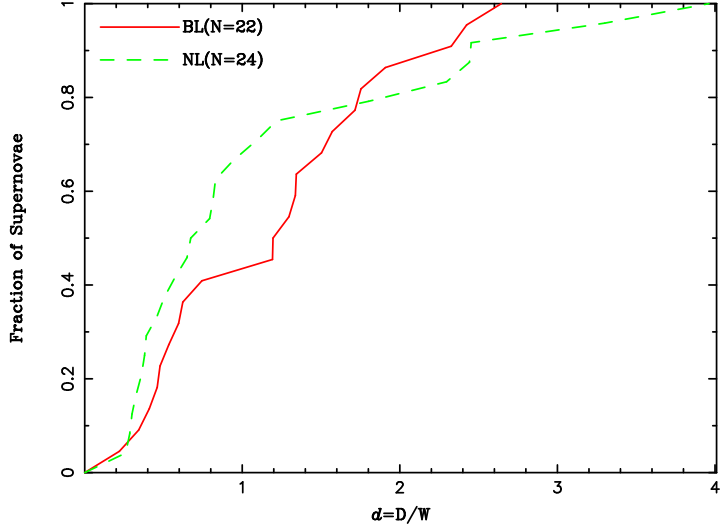}}
	\caption{The cumulative distributions of the relative distance from spiral arms for the BL (red solid) and NL (green dashed) SNe Ia, respectively, where D is the distance of a SN Ia from a spiral arm and W is the  semiwidth of the arm. The data are from \citet{KARAPETYAN22}.}\label{dgdis}
\end{figure*}

\subsection{Distance from spiral arms}\label{sect:3.8}
Based on the spiral density wave theory, stars form at shock fronts near the edges of spiral arms, as do the newly born SN progenitors (\citealt{LINCC64}). Then, the progenitors leaves their birth place until they arrive their explosion site. As a  result, the distance of a SN from the arm may be used to constrain the lifetime/population of the SN (\citealt{MIKHAILOVA07}; \citealt{ARAMYAN16}).  For example, it is found that SNe Ib/c locate closer the leading edges of the spiral arms than SNe II and the distance distribution of SNe Ia relative to the arms have broader wings than those of  SNe Ib/c  and SNe II (\citealt{MIKHAILOVA07}; \citealt{ARAMYAN16}). These facts indicate that  SNe Ibc have more massive progenitors than SNe II, and SNe Ia have less massive and older progenitors than CC SNe Ia. In particular, \citet{KARAPETYAN22} recently find that the $\Delta m_{\rm 15}{\rm (B)}$  value of the SNe Ia that locates on the arms are typically smaller than those of locating in the interarm region, i.e. on average, the arm SNe Ia are brighter than the interarm ones. The brightness difference between arm and inrerarm SNe Ia are believed from their different ages (\citet{KARAPETYAN22} ). In Fig.~\ref{dgdis}, we show the cumulative distributions of the relative distance from spiral arms for the BL and NL SNe Ia, respectively. Similar to the discussions in above sections, the two distributions are statistically indistinguishable, i.e. a K-S test gives a probability of 21.2\% that the two subsample are chosen from the same underlying mother sample.

\section{DISCUSSIONS AND CONCLUSIONS}\label{sect:4}
In this paper, we divide SNe Ia into two subclasses based on
their REW value, i.e. BL and NL SNe Ia, although the REW distribution of SNe Ia may perfectly fitted by one Gaussian. We want to investigate whether
or not there are systemical differences between BL and NL SNe Ia on
their environment in host galaxies. We find that comparing
with the NL SNe Ia, the distribution of $\Delta m_{\rm	15}{\rm (B)}$ for BL ones has a larger peak, and BL SNe Ia have a larger mean $\Delta m_{\rm
15}{\rm (B)}$. In addition, for the SNe Ia in spiral galaxies, there
seems a trend that the earlier the galaxy type, the larger the
number ratio of BL to NL SNe Ia, so do those in early type
galaxies. Moreover, on average, BL SNe Ia tend to occur in larger and brighter host galaxies then NL SNe Ia, but the distributions of the size and brightness of the host galaxies between BL and NL SNe Ia are indistinguishable. However, we do not find any significant difference between
BL and NL SNe Ia on their environments in their host galaxies, i.e. they have indistinguishable distributions of the fractional radial position, the fractional flux, the NCR pixel vale function and the fractional radial flux at their explosion sites, even similar distance distributions relative to spiral arms.  These indistinguishable distributions indicate that BL and NL SNe Ia share the same stellar populations.

In our sample, almost all HV SNe Ia belongs to the BL SNe Ia. \citet{WANGXF13} found that HV ones tend to occur in the metal-rich environment, which is confirmed further by \citet{PANYC15} and \citet{PANYC20}, who found that HV SNe Ia favor metal-rich massive host galaxies. In addition, \citet{WANGXF13} also found that  the mean value of the brightness of the HV SNe Ia is slightly dimmer than that of the NV ones. Our discovery that on average, BL SNe Ia are dimmer, and hosted by larger and brighter host galaxies than NL ones are consistent with previous studies.  The difference of the brightness between BL and NL SNe Ia could easily associate the higher
average $\Delta m_{\rm 15}{\rm (B)}$ value of BL SNe Ia with the
higher metallicity of their host galaxies, because it is well known
that the global metallicity of a galaxies increase with its
size/mass (\citealt{HENRY99}; \citealt{TREMONTI04}), and a high metallicity leads to a lower production of $^{\rm 56}$Ni and then a dimmer SN Ia (\citealt{TIMMES03})

In observations, the age and metallicity are two factors affecting the
properties of SNe Ia (\citealt{MENGXC15}). Many observations have
shown that the population age of SNe Ia is a key factor
affecting the luminosity of SNe Ia, and the dimmer SNe Ia may have a
wide age distribution (\citealt{GALLAGHER08}; \citealt{NEILL09};
\citealt{HOWEL09b}). Base on these observations, one may obtain a
conclusion that the average value and range of the luminosity of
SNe Ia decrease with their ages. On the other hand, both theory
and observation show that a higher metallicity leads to dimmer SNe
Ia (\citealt{TIMMES03}; \citealt{HOWEL09b}). In theory, about 25\% variation
of the peak luminosity of SNe Ia may result from the metallicity,
but observations show that metallicity may only contribute tot about 10\%
variation of the peak luminosity of SNe Ia (\citealt{TIMMES03};
\citealt{HOWEL09b}; \citealt{BRAVO10}). These could be derived from a kind of
degeneracy between the effect of age and metallicity on the brightness of SNe Ia since a high metallicity could also means a relatively younger SNe Ia and then a slightly brighter SNe Ia, whatever its progenitor is (\citealt{MENGXC11}; \citealt{MENGYANG12}). In observations, it is difficult to completely get rid of the effect of the age for the investigations on the correlation between the peak brightness and metallicity of SNe Ia.

\begin{figure}
\centerline{\includegraphics[angle=270,scale=.35]{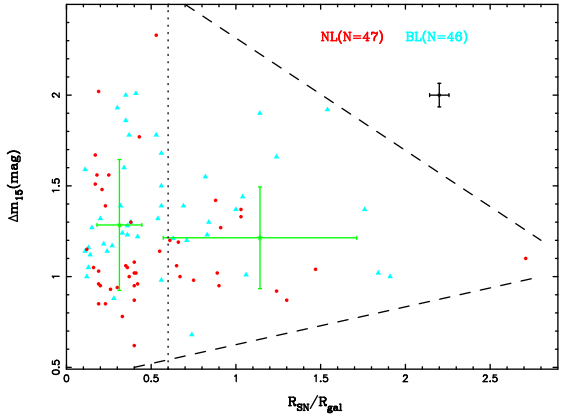}}
\caption{ $\Delta m_{\rm 15}{\rm (B)}$ vs.
$R_{\rm SN}/R_{\rm gal}$. The two green crosses show the average
values for SNe Ia in the inner and outer parts of their host
galaxies, respectively, and the lengthes of the bars present the
statical 1 $\sigma$ range. The black cross shows the typical error
of the points. The two dashed lines roughly show the upper and
lower boundaries in the ($\Delta m_{\rm 15}{\rm
	(B)}$ - $R_{\rm SN}/R_{\rm gal}$) plane. The data are from \citet{BLONDIN12} and \citet{WANGXF13}.}\label{rgaldeta}
\end{figure}

\begin{figure}
	\centerline{\includegraphics[angle=270,scale=.35]{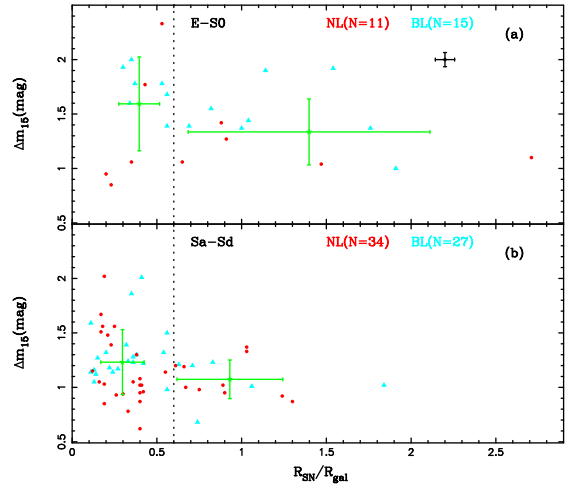}}
	\caption{Similar to Fig.~\ref{rgaldeta}, but for early and later type host galaxies, respectively.}\label{rgaldeta2}
\end{figure}

However, considering that  the environment may
reflect the age information of their populations to some extents 
(\citealt{FRUCHTER06}; \citealt{ANDERSON15}), the difference on the average $\Delta m_{\rm 15}{\rm (B)}$ value between BL and NL SNe Ia seems difficult to contribute
to the effect of age since there is not a significant difference on
the environment in their host galaxies. Especially, \citet{ANDERSON15b} divided their SN Ia sample into bright and dim subsamples to investigate the potential difference of  their environments within their host galaxies by pixel statistics method, but  no statistical significant difference is found. If the
metallicity is the main reason of the difference of the average
brightness between BL and NL SNe Ia, we would expect that on
average, the $\Delta m_{\rm 15}{\rm (B)}$ of the SNe Ia in the
inner part of their host galaxies is larger than that in the outer
part since on average, the metallicity for all kinds of galaxies
systematically decreases outward from their centers
(\citealt{HENRY99}). In Fig.~\ref{rgaldeta}, we show the relation
between $\Delta m_{\rm 15}{\rm (B)}$ and $R_{\rm SN}/R_{\rm gal}$.
It is clearly shown in the figure that the scatter of the $\Delta
m_{\rm 15}{\rm (B)}$ decrease with $R_{\rm SN}/R_{\rm gal}$, i.e.
the peak brightness of SNe Ia varies greater in the inner region
of their host galaxies than outer region, as found in
\citet{WANGLF97} and \citet{RIESS99}. Especially, the average
value of $\Delta m_{\rm 15}{\rm (B)}$ of SNe Ia in the inner region is
larger than that in the outer part by about 0.05 mag, as we
expected. At the same time, the 1 $\sigma$ statistical error of
$\Delta m_{\rm 15}{\rm (B)}$ of the inner SNe Ia is also large
than that of the outer ones by about 0.08 mag (\citealt{WANGLF97}; \citealt{RIESS99}). Nevertheless, it must be emphasized that such a difference
on the statistical error of $\Delta m_{\rm 15}{\rm (B)}$ is partly
contributed to the age effect.

In the above analysis, we mixed the SNe Ia in spiral and elliptical hosts, and
such a method can not get rid of the progenitor age effect. 
Here, we divided the host galaxies of SNe Ia into
two classes, i.e. E-S0 and Sa-Sd, which approximatively divides
SNe Ia into young and old classes. In Fig.~\ref{rgaldeta2}, we show the relation
between $\Delta m_{\rm 15}{\rm (B)}$ and $R_{\rm SN}/R_{\rm gal}$ for the above two classes. The figure shows
that the difference of the average $\Delta m_{\rm 15}{\rm (B)}$
between inner and outer SNe Ia is even enlarged compared with that
in Fig.~\ref{rgaldeta}, and the differences (the average  $\Delta m_{\rm 15}{\rm (B)}$ differences between the inner and outer part SNe Ia are 0.25 mag and 0.16 mag for E-S0 and Sa-Sd classes, respectively) are larger enough to explain the average brightness difference between BL and NL SNe Ia shown in Fig.~\ref{rrdeta}. The results in Figs.~\ref{rgaldeta} and \ref{rgaldeta2} indicate that there indeed exists a kind of degeneracy for the effect between the metallicity and progenitor age on the peak brightness of SNe Ia.

The global metallicity of a galaxies increase with its
size/mass, and based on the mass - metallicity relation, one may obtain the global gas-phase metallicity (\citealt{HENRY99}; \citealt{TREMONTI04}; \citealt{SULLIVAN10}). However, such a global mean metallicity could erase  the intrinsic information of the population at the explosion site, because the global parameter of a host galaxy may only
represent the average information of the stellar population in the
host galaxy  (\citealt{MENGXC19}). This could be the reason of the fact that  on average, BL SNe Ia tend to occur in larger and brighter host
galaxies then NL SNe Ia, with the statistically indistinguishable distribution of the brightness and size of the host galaxies for BL and NL SNe Ia. The local gas-phase metallicity around the explosion site of a SN Ia is perfect tracer of the metallicity of the SN Ia (\citealt{SULLIVAN10}; \citealt{MORENORAYA16}).  \citet{MORENORAYA16} found that the absolute magnitude and color of SNe Ia are well correlated with the local gas-phase oxygen abundance. If the difference of the brightness between BL and NL SNe Ia is derived from their different metallicities, we would expect that the  local gas-phase metallicity for BL SNe Ia is statistically higher than those for NL ones.  Based on the data in \citet{MORENORAYA16}, as expected, we notice that the local gas-phase average value of 12 + log(O/H) for BL SNe Ia is $8.567\pm0.081$, higher than $8.389\pm0.138$ for NL ones. A K-S test shows that the probability that the two subsamples are from the same mother sample is only 4.9\%. As a conclusion, the metallicity for BL SNe Ia is statistically larger than that for NL ones. This is consistent with the fact that BL SNe Ia is more possible to occur in elliptical galaxies (Fig.~\ref{rrhost}). The results here indicate that the progenitor metallicity is a key parameter to affect the  photometric and spectroscopic diversities of SNe Ia (\citealt{MAGUIRE12}; \citealt{WANGXF13}; \citealt{TAUBENBERGER17}).

As far as the progenitor systems are concerned, different pieces of evidence  usually favor different progenitor models. For instance, \citet{WANGXF19} found that  HV SNe Ia have a significant excess flux in blue light 60-100 days past maximum, while NV SNe do not show such an excess blue flux, which indicates that  HV SNe Ia are likely from SD systems, while the NV ones may arise from DD systems. However, considering the relation between photospheric velocity and the velocity shift of later-time [Fe II] lines, which indicates an asymmetric explosion for SNe Ia (\citealt{MAEDA10}; \citealt{MAGUIRE18}), \citet{LIWX21} suggest the HV and a portion of NV object are from double-detonation model and their different photospheric velocities are from different view angle (see also \citealt{PANYC20}).  Nevertheless, considering the correlation between the progenitor population and the characteristics of the high velocity feature (HVF) of Ca II IR3 absorption feature around maximum light, \citet{MENGXC19} suggests that the SD model has a potential to simultaneously explain all the characteristics of HVFs observed. However, we do not notice any significant difference for the environment between BL and NL SNe Ia, although BL SNe Ia are statistically dimmer than NL ones, which seems not to provide a constraint to the progenitors of SNe Ia. 

In physics, the REW of the  Si II 635.5 nm absorption feature of a SN Ia around maximum light represents the velocity interval of the silicon layer in supernova ejecta to a great extent. \citet{MENGXC17} found that, if taking the REW of the  Si II 635.5 nm absorption feature of SNe Ia as the parameter describing the SNe Ia, all kinds of SNe Ia follow a universal polarization sequence, which implies that all SNe Ia could share the same explosion mechanism, no matter what their progenitor systems are. If the REW of the  Si II 635.5 nm absorption feature of a SN Ia reflects the explosion mechanism of SNe Ia to some extent, our results that there is not difference on the progenitor population of SNe Ia with different  REW of the  Si II 635.5 nm line imply that no matter what the progenitor systems are, the explosion mechanism of SNe Ia could be similar or the same. The conclusion here is consistent with that suggested by \citet{MENGXC17}. 

In summary, based on the  REW of the  Si II 635.5 nm line around the maximum light, we divide SNe Ia into BL and NL subsamples, and find that the BL SNe Ia are statistically dimmer than the NL ones.  On average, the BL SNe Ia are hosted by brighter and larger host galaxy, and the number ratio of BL to NL SNe Ia increases with their host galaxies becoming early.  These facts are possibly derived from the effect of metallicity, i.e. the BL SNe Ia have a higher metallicity than the NL ones.  However, according to pixel statistics study, we do not find any potential difference of the explosion environment around the explosion sites between BL and NL SNe Ia. These results indicate that all SNe Ia could share the same/similar explosion mechanism, which does not depend on their stellar population.

\section*{Acknowledgments}
This work was supported by the National Key R\&D Program of China with No. 2021YFA1600403, and the NSFC (Nos. 11973080 and 11733008). XM acknowledges science research grants from the China Manned Space Project, no. CMS - CSST - 2021 - B07. XM acknowledges support from the Yunnan Ten Thousand Talents Plan - Young \& Elite Talents Project, and the CAS `Light of West China' Program. J.Z. is supported NSFC (grants 12173082, 11773067 ),  the CAS `Light of West China' Program and the Ten Thousand Talents Program of Yunnan for Top-notch Young Talents.  We acknowledge the usage of the HyperLeda database (http://leda.univ-lyon1.fr)

\end{document}